\documentclass[twocolumn,secnumarabic,floatfix,amssymb,amsmath,nofootinbib,tightenlines,showpacs,superscriptaddress,aps,prb]{revtex4}
\usepackage{amsfonts}
\usepackage[colorlinks=true,linkcolor=blue,pagecolor=blue,filecolor=blue,menucolor=yellow,urlcolor=blue,citecolor=blue,anchorcolor=blue]{hyperref}%
\usepackage{graphicx}
\usepackage{dcolumn}
\usepackage{times}

\newcommand{\e}[1]{\mathrm{e}^{#1}}

\newcommand{\ie}{\textit{i.e. }}
\newcommand{\eg}{\textit{e.g. }}
\newcommand{\etal}{\emph{et al.}}
\newcommand{\pr}{Phys. Rev. }
\def\i{\mathrm{i}}

\begin{document}

\title{Superconducting Phase Transistor in Diffusive Four-terminal Ferromagnetic Josephson Junctions}

\author{Mohammad Alidoust }
\email{phymalidoust@gmail.com} \affiliation{Department of Physics,
Norwegian University of Science and Technology, N-7491 Trondheim,
Norway}
\author{Granville Sewell}
\email{sewell@utep.edu} \affiliation{Mathematics Department,
University of Texas El Paso, El Paso, TX 79968, USA}

\author{Jacob Linder}
\email{jacob.linder@ntnu.no} \affiliation{Department of Physics,
Norwegian University of Science and Technology, N-7491 Trondheim,
Norway}
\date{\today}

\begin{abstract}
We study diffusive magnetic Josephson junctions with four
superconducting terminals in the weak proximity limit where the
leads are arranged in cross form. Employing the linearized
Keldysh-Usadel technique, the anomalous Green's function and
Josephson current are analytically obtained based on a
quasiclassical theory using the Fourier series method. The derived
results may be reduced to non-magnetic junctions by setting the
exchange field equal to zero. We find that increments of the
magnetic barrier thickness may cause a reversal of the supercurrent
direction flowing into some of the leads, whereas the direction of
current-flow remains invariant at the others. The reversal direction
can be switched by tuning the perpendicular superconducting phases.
In the non-magnetic case, we find that the supercurrent flowing
between the leads in one direction can be tuned by changing the
superconducting phase difference in the perpendicular direction.
These findings suggest the possibility of constructing a nano-scale
superconducting phase transistor whose core element consists of the
proposed four-terminal Josephson junction with rich switching
aspects.
\end{abstract}


\maketitle

\section{introduction}
When a weak link is established between two superconductors, a
gradient in the superconducting
phases can drive a supercurrent through the system. This Josephson effect \cite{cite:josephson,cite:yason,cite:shapiro} and the associated current-phase relation in
weak links has been investigated extensively in previous literature,
see for example the comprehensive reviews Refs.
\onlinecite{cite:likharev} and \onlinecite{cite:golubov} (see also
Refs. \onlinecite{cite:buzdin2} and \onlinecite{cite:bergeret} for
magnetic Josephson junctions).

The proximity effect between superconductors
and normal diffusive metals was first studied by W.L. McMillan in
1965 \cite{cite:mcmillam}. It is known that the electronic
properties of a normal metal become altered when placed in proximity
to a host superconductor. For instance, the electronic spectrum of
the normal metal connected to a superconductor exhibits a minigap
\cite{cite:mcmillam,cite:belzig,cite:hammer1,cite:hammer2,
cite:sueur_prl_10, cite:alidoust_prb_10}. Very recently, the key
properties of density of states (DOS) of a sandwiched normal metal
between superconductors were employed in an experiment for producing
a superconducting quantum interference proximity transistor (SQUIPT)
\cite{cite:giazotto}. Moreover,
superconductor-normal metal-superconductor (S/N/S) Josephson
junctions have been studied under non-equilibrium conditions where
two additional normal leads are connected to the sandwiched normal
layer. It has been demonstrated that this type of S/N/S Josephson
junctions is able to produce a $\pi$-junction depending on the
applied voltage to the normal sandwiched layer
\cite{cite:crosser,cite:Baselmans}. Such $\pi$-junctions may also be
observed in three terminal junctions \cite{cite:crosser,cite:Huang}.

The proximity-induced interplay between superconductivity and
ferromagnetism in hybrid structures is also known to establish intriguing
physical phenomena. The wavefunction describing the leakage of
Cooper pairs inside a ferromagnet oscillates in a damped fashion.
One of the most interesting phenomena in the proximity of
ferromagnetism and superconductivity is 0-$\pi$ transition which may
occur in superconductor-ferromagnet-superconductor (S/F/S)
junctions \cite{cite:buzdin1,cite:buzdin2,cite:Garifullin,cite:Sidorenko,cite:Jiang}. The transition
usually occurs over a narrow length $\xi_F=\sqrt{D_F/h}$ in which
$D_F$ and $h$ represent the diffusion constant and the exchange
field of the sandwiched ferromagnetic layer, respectively. At this
crossover point, the minimum energy of junction is switched between
zero and $\pi$-superconducting phase difference by changing the
energy scales of the system such as Thouless energy, exchange field
and temperature. Also it has been demonstrated
that the spin-flip scattering may render the junction energy minimum
from 0 to $\pi$ \cite{cite:ryazanov1,cite:buzdin2,cite:buzdin4,
cite:linder_prb_08} and that the supercurrent itself may become spin-polarized if the magnetization texture is inhomogeneous \cite{cite:spin_josephson}.

So far in the literature, the main emphasis has mostly been on
one-dimensional systems where two superconductors are coupled via
\eg a constriction or diffusive metal. On the other hand, the
interplay between multiple superconducting terminals
\cite{cite:crosser} in a Josephson junction would require an
extension to higher dimensions \cite{cite:malek1,cite:amin2}. This
in turn complicates the analytical treatment of the system, and one
is usually forced to resort to numerical means within the diffusive
regime \cite{cite:bergeret2}. It would therefore be of interest to
clarify how the transport characteristics of a diffusive
ferromagnetic Josephson junction is influenced by the presence of
multiple superconducting phase differences, and also to provide an
analytical framework for studying such phenomena. Multi-terminal
Josephson point contacts had intensively been investigated (both AC
and DC characteristics) using the Ginzburg-Landau theory
\cite{cite:omel1,cite:omel2,cite:omel3} and was followed by studying
the four-terminal S/N/S Josephson junctions in the clean limit via
the Eilenberger equations \cite{cite:malek1,cite:malek2,cite:amin}.
Interesting phenomena such as phase dragging (the production of
phase difference between two terminals by means of phase variation
between other terminals), magnetic flux transfer and  bistable
states were found due to non-local coupling and additional degrees
of freedom in such classes of Josephson junctions
\cite{cite:malek1,cite:malek2,cite:omel1}. Such point contacts also
have been fabricated and intensively studied in experiments
\cite{cite:omel2}.

Motivated by this, we consider in this paper a diffusive Josephson
junction with four superconducting leads where are arranged in a
cruciate form and study the supercurrent flowing in this junction.
The superconducting leads are separated by a metal that may or may
not be ferromagnetic. We use the quasiclassical Usadel equations in
the diffusive regime and formulate the current-phase relation as a
function of all the available parameters in the system such as
superconducting phases in the magnetic junction. We recover the
results of Refs. \onlinecite{cite:malek1} and
\onlinecite{cite:malek2} obtained in the clean S/N/S junctions:
namely, when the dimensions $L$ (length) and $W$ (width) of the
sandwiched metal are comparable to each other, \ie $L\simeq W$, the
standard sinusoidal supercurrent is strongly modified by all the
condensate phases. We also use a phenomenological Ginzburg-Landau
theory to confirm our analytical expressions obtained via the
quasiclassical framework. In particular, we demonstrate that the
Josephson current flowing between leads along one axis may be tuned
via the superconducting phase gradient in the perpendicular
direction.


Moreover, we find that increments of the magnetic barrier thickness
may cause a reversal of the supercurrent direction flowing into some
of the leads, whereas the direction of current-flow remains
invariant at the others. These findings are suggestive in terms of
designing a nano-scale superconducting phase transistor where
current switching effects in one direction is possible by variation
of macroscopic superconducting phase in the perpendicular direction
as has also been pointed out in Ref. \onlinecite{cite:amin} and
\onlinecite{cite:amin2} for ballistic contacts.


The paper is organized as follows. In Sec. \ref{sec:Theory} we
present our main analytical findings. In Subsect. \ref{subsec:GF}
the basic equations of the quasiclassical method are presented and
in Subsect. \ref{subsec:GF_current} the cruciate Josephson junction
is studied analytically via the Green's function method. We
formulate the current-phase relation as a function of the four
superconducting phases for a magnetic Josephson junction. In
Subsect. \ref{subsec:GL} we confirm our results and findings via a
macroscopic Ginzburg-Landau theory. In Sec. \ref{sec:SFS} we employ
a 'Jacobi' numerical method \cite{cite:alidoust} (which shall be
explained in detail) and investigate the behavior of the
supercurrent which confirms our analytical derived expressions in
Subsect. \ref{subsec:GF_current} and their dependencies on the
superconducting U(1) phases, also the behavior of junction is
analyzed in more detail. Sec. \ref{sec:SFS} is devoted to the study
of the supercurrent behavior in S/F/S four-terminal junctions as a
function of ferromagnetic barrier thickness. Concluding remarks are
finally given in Sec. \ref{sec:Conclusions}.

\section{Theory and analytical discussions}\label{sec:Theory}
\begin{figure}[t!]
\includegraphics[width=6.5cm,height=4.cm]{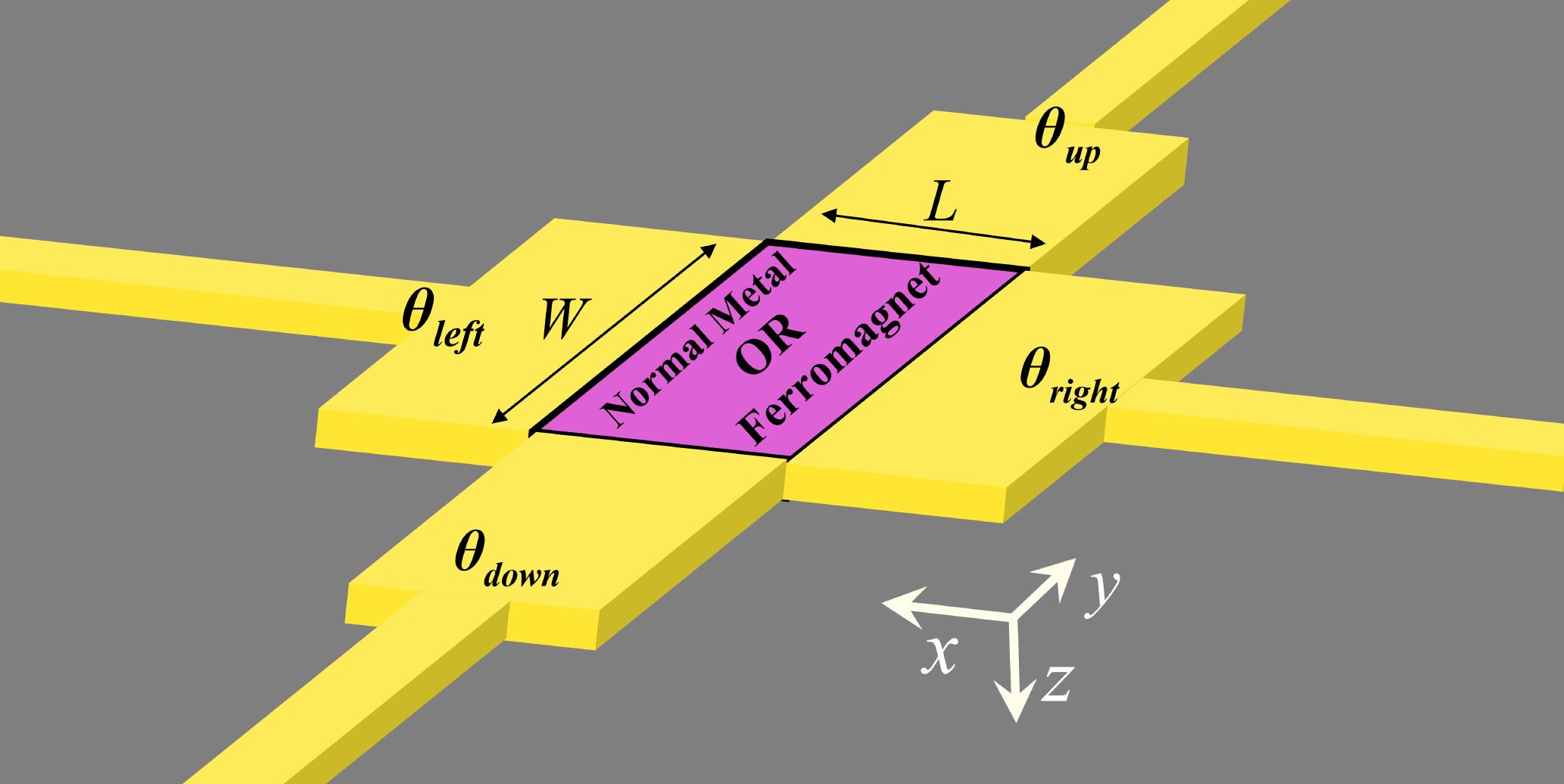}
\caption{\label{fig:model} Experimental schematic setup of the
cruciate Josephson junction. The junction is assumed to lie in the
$xy$ plane with interfaces located at $x=0, L$ and $y=0, W$. The
four spin-singlet superconductors have different superconducting
phases: $\theta_{\text{up}}, \theta_{\text{down}},
\theta_{\text{left}}$ and $\theta_{\text{right}}$. Exchange field
$\textbf{h}$, is assumed to be oriented in the $z$ direction
perpendicular to the sandwiched layer plane.}
\end{figure}

We consider four superconducting leads coupled via a ferromagnetic
or normal diffusive metal. As in Fig. \ref{fig:model}, the
nano-scale diffusive metal is assumed to be located in the $xy$
plane, where $x\in[0,L]$ and $y\in[0,W]$. The four superconducting
terminals are assumed to have equal magnitudes for the gap $\Delta$
and are connected to each edge of the diffusive strip. The
suppression of the pair potential is neglected near interfaces due
to a low interface transparency and the superconducting phases are
assumed to be different in each of the four terminals:
$\theta_{\text{up}}, \theta_{\text{down}}, \theta_{\text{left}}$ and
$\theta_{\text{right}}$. One may expect that superconducting
correlations inside the system interfere, resulting in a quite
complicated coherent system. The S/F/S system is studied in the
diffusive limit and current-phase relationship is obtained at each
terminal similar to clean S/N/S four-terminal junctions
\cite{cite:malek1,cite:malek2}. In our approach, we start with a
magnetic four-terminal Josephson junction and derive our analytical
results for the magnetic system. We then may achieve the
non-magnetic Josephson junction characteristics by setting the
magnetic exchange field $h$ equal to zero.

\subsection{Microscopic Green's function approach}\label{subsec:GF}
In this subsection, we present basic equations of the quasiclassical
Keldysh-Usadel method. In order to study the transport properties of
the proposed four-terminal device, we employ the quasiclassical
method. In the diffusive regime, due to the existence of strong
scattering sources, quasiparticles' momentums are integrated over
all directions in space. In this case, the Eilenberger equations
reduce to the Usadel equations \cite{cite:usadel}. Under equilibrium
conditions, the system under consideration can be described by a
$4\times 4$ matrix propagator in Nambu space: the retarded Green's
function $G^{R}$. The total Green's function describing the system
compactly reads \cite{cite:mortenthesis}:
\begin{equation}\label{eq:greenfunc}
    \hat{G}(R,\varepsilon,T)=\left(\begin{array}{cc}
              G^{A} & G^{K} \\
              \mathbf{0} & G^{R}
            \end{array}\right),\;G^{R}=\left(\begin{array}{cc}
              g^{R} & f^{R} \\
              -\tilde{f} & -\tilde{g}
            \end{array}\right),
\end{equation}
where the meaning of the $\tilde{...}$-operation depends on the
notation adopted. In our notation, it denotes complex conjugation
and a change in sign for the energy argument. The advanced and
Keldysh blocks are made from retarded block by
$G^{A}=-(\tau_3G^R\tau_3)^{\dag}$ and
$G^{K}=\tanh(\beta\varepsilon)(G^{R}-G^{A})$ in which $\tau_3$ is
the Pauli matrix and $\beta=k_BT/2$. In the presence of exchange
energy $\textbf{h}=(h_x,h_y,h_z)$ inside the ferromagnetic layer,
the Usadel equation can be give by;
\begin{align}\label{Advanced}
D[\hat{\partial},\hat{G}[\hat{\partial},\hat{G}]]+i[ \varepsilon
\hat{\rho}_{3}+
\text{diag}[\textbf{h}\cdot\underline{\sigma},(\textbf{h}\cdot\underline{\sigma})^{\tau}],\hat{G}]=0,
\end{align}
where $\hat{\rho}_{3}$ and $\underline{\sigma}$ are $4\times 4$ and
$2\times 2$ Pauli matrixes, respectively. Here  $D$ is diffusive
constant of the sandwiched medium. Also, $\varepsilon$ is the
quasiparticles' energy which is measured from Fermi surface.

The so-called weak proximity regime occurs in the case of very low
transparent interfaces or for temperatures near to the critical
temperature of the superconducting leads. The superconducting
correlations leak into the ferromagnetic region weakly and so the
normal and anomalous Green's functions can be approximated by
 $\underline{\text{g}}\simeq \underline{1}$
and $\underline{f}\ll \underline{1}$, respectively. In this limit
one can linearize the Usadel equation which yields a set of
uncoupled complex boundary value partial differential equations. The
energy representation is used in this paper, however, one may reach
the Matsubara representation by replacing $\varepsilon\rightarrow
i\omega_n$, where $\omega_n=(2n+1)\pi k_BT$ are Matsubara
frequencies. For the sake of simplicity, a uniform exchange field
for the ferromagnetic layer is considered throughout the paper \ie
$\textbf{h}=(0,0,h_z=h)$. In the weak proximity regime that
mentioned above, the Green's function read \cite{cite:linder_prb_08}
\begin{align}
\hat{G}^{R}\approx\begin{pmatrix}
\underline{1} & \underline{f}^{R}\\
-\underline{\tilde{f}}^{R} & -\underline{1}\\
\end{pmatrix},
\end{align}
in fact, we have expanded the Green's function around the bulk
solution $\hat{G}_{0}$ as $\hat{G}\simeq\hat{G}_{0}+\hat{f}$, where
$\hat{G}_{0}=\text{diag(\underline{1},-\underline{1})}$
\cite{cite:bergeret}. The retarded Green's function now can be given
by;
\begin{align}\label{Advanced Gree}
\hat{G}^{R}=\begin{pmatrix}
1 & 0 & 0 & f^{R}_{+}(\varepsilon) \\
0 & 1  &f^{R}_{-}(\varepsilon)  & 0 \\
 0  & [-f^{R}_{+}(-\varepsilon)]^{\ast}   & -1  &  0  \\
[-f^{R}_{-}(-\varepsilon)]^{\ast}& 0  & 0  &  -1  \\
\end{pmatrix}.
\end{align}
If we assume that the exchange field is uniform throughout the
sample and is oriented in the $z$ direction, so the Usadel equations
reduce to two dimensional form as belows:
\begin{eqnarray}\label{eq:Linearized Usadel Eq.1}
\partial_{x}^{2}f^{R}_{\pm}(-\varepsilon)+\partial_{y}^{2}f^{R}_{\pm}(-\varepsilon)-\frac{2i(\varepsilon\mp
h)}{D}f^{R}_{\pm}(-\varepsilon)=0,
\end{eqnarray}
\begin{eqnarray}\label{eq:Linearized Usadel Eq.2}
\partial_{x}^{2}[f^{R}_{\pm}(\varepsilon)]^{*}+\partial_{y}^{2}[f^{R}_{\pm}(\varepsilon)]^{*}-\frac{2i(\varepsilon\pm
h)}{D}[f^{R}_{\pm}(\varepsilon)]^{*}=0.
\end{eqnarray}
We employ the Kupriyanov-Lukichev boundary conditions at F/S
interfaces \cite{cite:zaitsev} and control their opacities using a
parameter $\zeta$ that depends on the resistance of the interface
and the diffusive normal region;
\begin{equation}\label{eq:bc}
    \zeta(\hat{G}\hat{\partial}\hat{G})\cdot\hat{\boldsymbol{n}}=[\hat{G}_{\text{BCS}}(\theta),\hat{G}],
\end{equation}
where $\hat{\boldsymbol{n}}$ is a unit vector denoting the
perpendicular direction to an interface. The bulk solution ,
$\hat{G}_{\text{BCS}}$ for a $s$-wave superconductor is
\cite{cite:mortenthesis};
\begin{align}
\hat{G}^{R}_{\text{BCS}}(\theta)=\left(
                                  \begin{array}{cc}
                                    \mathbf{1}\cosh(\vartheta(\varepsilon)) & i\tau_2\sinh(\vartheta(\varepsilon))e^{i\theta} \\
                                    i\tau_2\sinh(\vartheta(\varepsilon))e^{-i\theta} & -\mathbf{1}\cosh(\vartheta(\varepsilon)) \\
                                  \end{array}
                                \right),
\end{align}
\begin{equation}
\nonumber
\vartheta(\varepsilon)=\text{arctanh}(\frac{\mid\Delta\mid}{\varepsilon}),
\end{equation}
\begin{eqnarray}
&&\nonumber s(\varepsilon)\equiv\sinh(\vartheta(\varepsilon))e^{i\theta}=\\&&\nonumber-\Delta\left\{\frac{\text{sgn}(\varepsilon)}{\sqrt{\varepsilon^2-\Delta^2}}\Theta(\varepsilon^2-\Delta^2)-\frac{i}{\sqrt{\Delta^2-\varepsilon^2}}\Theta(\Delta^2-\varepsilon^2)\right\}\\
&&\nonumber
c(\varepsilon)\equiv\cosh(\vartheta(\varepsilon))=\\&&\nonumber\frac{\mid\varepsilon\mid}{\sqrt{\varepsilon^2-\Delta^2}}\Theta(\varepsilon^2-\Delta^2)-\frac{i\varepsilon}{\sqrt{\Delta^2-\varepsilon^2}}\Theta(\Delta^2-\varepsilon^2).
\end{eqnarray}
$\Delta$ is superconducting gap in the $s$-wave superconductors and
the Heaviside step-function is denoted by $\Theta(\varepsilon)$. In
this paper, we have defined $\theta_u$, $\theta_d$, $\theta_l$,
$\theta_r$ as the condensate phases in the up, down, left and right
superconductor leads, respectively. If we now open up the compacted
boundary conditions Eq. (\ref{eq:bc}) at left F/S interface for
instance, $x=0$, we reach at;
\begin{eqnarray}\label{eq:B.C. 1}
&&\nonumber(\zeta\partial_{x} -c^*(\varepsilon)
)f^{R}_{\pm}(-\varepsilon)=\pm s^*(\varepsilon)e^{i\theta_{l}}\\&&
(\zeta\partial_{x} -c^*(\varepsilon)
)[f^{R}_{\pm}(\varepsilon)]^{*}=\mp
s^*(\varepsilon)e^{-i\theta_{l}},
\end{eqnarray}
 and at $x=L$
\begin{eqnarray}\label{eq:B.C. 2}
&&\nonumber(\zeta\partial_{x} +c^*(\varepsilon)
)f^{R}_{\pm}(-\varepsilon)=\mp s^*(\varepsilon)e^{i\theta_{r}}\\&&
(\zeta\partial_{x}+c^*(\varepsilon)
)[f^{R}_{\pm}(\varepsilon)]^{*}=\pm
s^*(\varepsilon)e^{-i\theta_{r}}.
\end{eqnarray}
Also at $y=0$
\begin{eqnarray}\label{eq:B.C. 3}
&&\nonumber(\zeta\partial_{y} -c^*(\varepsilon)
)f^{R}_{\pm}(-\varepsilon)=\pm s^*(\varepsilon)e^{i\theta_{d}}\\&&
(\zeta\partial_{y} -c^*(\varepsilon)
)[f^{R}_{\pm}(\varepsilon)]^{*}=\mp
s^*(\varepsilon)e^{-i\theta_{d}},
\end{eqnarray}
 and at $y=W$ the boundary condition takes the below form
\begin{eqnarray}\label{eq:B.C. 4}
&&\nonumber(\zeta\partial_{y} +c^*(\varepsilon)
)f^{R}_{\pm}(-\varepsilon)=\mp s^*(\varepsilon)e^{i\theta_{u}}\\&&
(\zeta\partial_{y}+c^*(\varepsilon)
)[f^{R}_{\pm}(\varepsilon)]^{*}=\pm
s^*(\varepsilon)e^{-i\theta_{u}}.
\end{eqnarray}

In the equilibrium conditions, the current density vector is given
by Keldysh block as
\begin{equation}\label{eq:currentdensity}
{\mathbf{J}}\text{(}\mathbf{R}\text{)}=J_{0}\int
d\varepsilon\text{Tr}\{\rho_{3}(\hat{G}[\hat{\partial},\hat{G}])^{K}\}
\end{equation}
here $J_{0}$ is a normalization constant. The current density vector determines the direction
and amplitude of current density inside the sandwiched layer as a
function of coordinates. If we substitute the total Green's function
Eq. (\ref{eq:greenfunc}) into the current density relation namely,
Eq. (\ref{eq:currentdensity}) we arrive at:
\begin{eqnarray}\label{eq:current}
&&\nonumber\mathbf{J}(\mathbf{R})=J_{0}\int_{-\infty}^{\infty}d\varepsilon\tanh(\varepsilon
\beta)\left\{
f^{R}_{-}(-\varepsilon)\vec{\nabla}[f^{R}_{+}(\varepsilon)]^{*}\right.\\&&\nonumber+
f^{R}_{+}(-\varepsilon)\vec{\nabla}[f^{R}_{-}(\varepsilon)]^{*}-
f^{R}_{+}(\varepsilon)\vec{\nabla}[f^{R}_{-}(-\varepsilon)]^{*}-f^{R}_{-}(\varepsilon)\\&&\nonumber\vec{\nabla}[f^{R}_{+}(-\varepsilon)]^{*}+
[f^{R}_{-}(-\varepsilon)]^{*}\vec{\nabla}f^{R}_{+}(\varepsilon)+[f^{R}_{+}(-\varepsilon)]^{*}\vec{\nabla}f^{R}_{-}(\varepsilon)\\&&\left.-
[f^{R}_{+}(\varepsilon)]^{*}\vec{\nabla}f^{R}_{-}(-\varepsilon)-[f^{R}_{-}(\varepsilon)]^{*}\vec{\nabla}f^{R}_{+}(-\varepsilon)\right\}.
\end{eqnarray}
To obtain total supercurrent flowing through the junction, for
example at right superconducting gate, one needs to perform an
integration of Eq. (\ref{eq:currentdensity}) over the $y$ coordinate
, $I\text{(}\phi\text{)}=I_{0}\int\int
dyd\varepsilon\text{Tr}\{\rho_{3}(\check{\text{g}}[\hat{\partial},\check{\text{g}}])^{K}\}$.

At this point it suffices that Eqs. (\ref{eq:Linearized Usadel Eq.1}) be
solved together with appropriate boundary conditions (\ie Eqs.
(\ref{eq:B.C. 1}), (\ref{eq:B.C. 2}), (\ref{eq:B.C. 3}) and
(\ref{eq:B.C. 4})) in order to capture the transport characteristics of
the present class of Josephson junctions in the diffusive limit.

\subsection{Analytical microscopic
discussions}\label{subsec:GF_current}
In this subsection we derive
explicit analytical expressions describing the supercurrent at
each superconducting terminal. To this end, we consider the weak
proximity limit of diffusive regime where the Keldysh-Usadel method
yields a set of uncoupled complex elliptic partial differential
equations. The simplified Usadel equations and corresponding
boundary conditions are give by Eqs. (\ref{eq:Linearized Usadel
Eq.1}), (\ref{eq:Linearized Usadel Eq.2}), (\ref{eq:B.C. 1}),
(\ref{eq:B.C. 2}), (\ref{eq:B.C. 3}) and (\ref{eq:B.C. 1}). For
simplicity in our analytical calculations we exclude first-order
terms of the anomalous Green's function in the Kupryianov-Lukichev
boundary conditions, Eq. (\ref{eq:bc}). We use the Fourier series
method in the presence of non-homogenous boundary conditions and
obtain analytical solutions for the Usadel equations. The method
leads a somewhat lengthy solutions, for instance one of the
anomalous components of Green's function namely,
$f_{+}^{R}(\varepsilon)$ after long calculations is given by Eq.
(\ref{eq:fp});
\begin{widetext}
\begin{eqnarray}\label{eq:fp}
  \nonumber f_{+}^{R}(\varepsilon)&&=-\left\{\frac{\Delta\text{sgn}(\varepsilon)}{\sqrt{\varepsilon^2-\Delta^2}}\Theta(\varepsilon^2-\Delta^2)-\frac{i\Delta}{\sqrt{\Delta^2-\varepsilon^2}}\Theta(\Delta^2-\varepsilon^2)\right\}\left\{\frac{e^{i\theta_l}}{L\zeta}(x-\frac{x^2}{2L}+\frac{D}{2iL(\varepsilon+h)}-\frac{L}{3}-\right.\\&&\nonumber\sum_{k=1}^{\infty}\frac{4iL(\varepsilon+h)\cos(\frac{k\pi
x}{L})}{k^2\pi^2(Dk^2\pi^2/L^2-2i(\varepsilon+h))})
-\frac{e^{i\theta_r}}{L\zeta}(\frac{x^2}{2L}-\frac{D}{2iL(\varepsilon+h)}-\frac{L}{6}+\sum_{k=1}^{\infty}\frac{4iL(\varepsilon+h)(-1)^k\cos(\frac{k\pi
x}{L})}{k^2\pi^2(Dk^2\pi^2/L^2-2i(\varepsilon+h))})\\&&\nonumber+
\frac{e^{i\theta_d}}{W\zeta}(y-\frac{y^2}{2W}+\frac{D}{2iW(\varepsilon+h)}-\frac{W}{3}-\sum_{l=1}^{\infty}\frac{4iW(\varepsilon+h)\cos(\frac{l\pi
y}{W})}{l^2\pi^2(Dl^2\pi^2/W^2)-2i(\varepsilon+h)})-\frac{e^{i\theta_u}}{W\zeta}(\frac{y^2}{2W}-\frac{D}{2iW(\varepsilon+h)}\\&&\left.-\frac{W}{6}+\sum_{l=1}^{\infty}\frac{4iW(\varepsilon+h)(-1)^l\cos(\frac{l\pi
y}{W})}{l^2\pi^2(Dl^2\pi^2/W^2-2i(\varepsilon+h))})\right\}.
 \end{eqnarray}
 \end{widetext}
The length and width of the ferromagnetic region sandwiched between
the superconductors are denoted by $L$ and $W$. As can be seen, the
anomalous component of the retarded Green's function depends on all
four condensation phases, which in turn leads to an interference
between these superconducting phases in the Josephson current. In
Eq. (\ref{eq:current}) there are 8 different terms of anomalous
component of Green's function involved the supercurrent relation.
Therefore, one must find 8 similar solutions as Eq. (\ref{eq:fp})
for other terms and substitute them into the supercurrent relation
Eq. (\ref{eq:current}) in order to obtain the supercurrent at
one terminal. To obtain analytical solutions for the total supercurrent
flowing at the other superconducting terminals, one must repeat the
latter described process. We have done so and arrived at the
analytical expressions describing the supercurrent in the system as
follows. Supercurrent at $x=0, L$ terminals are obtained as

\begin{eqnarray}\label{eq:I_x0}
&&\nonumber
\frac{I_x(x=0)}{I_{0}}=\int_{-\infty}^{\infty}\frac{d\varepsilon}{\Delta_0}\frac{\Delta^2\tanh(\beta\varepsilon)}{\Delta^2-\varepsilon^2}\sum_{\sigma=\pm}\left\{(\frac{WD}{L^3\zeta^2(\varepsilon+\sigma
h)}\right.\\&&\nonumber\left.+\frac{8WD}{L^3\zeta}\sum_{k=1}^{\infty}\frac{(-1)^k(\varepsilon+\sigma
h)}{D^2k^4\pi^4/L^4+4(\varepsilon+\sigma
h)^2})\sin(\theta_l-\theta_r)+\right.\\&&\left.\frac{D\sin(\theta_l-\theta_u)}{LW\zeta^2(\varepsilon+\sigma
h)}+\frac{D\sin(\theta_l-\theta_d)}{LW\zeta^2(\varepsilon+\sigma
h)}\right\}
\end{eqnarray}
\begin{eqnarray}\label{eq:I_xL}
&&\nonumber
\frac{I_x(x=L)}{I_{0}}=\int_{-\infty}^{\infty}\frac{d\varepsilon}{\Delta_0}\frac{\Delta^2\tanh(\beta\varepsilon)}{\Delta^2-\varepsilon^2}\sum_{\sigma=\pm}\left\{(\frac{WD}{L^3\zeta^2(\varepsilon+\sigma
h)}\right.\\&&\nonumber\left.+\frac{8WD}{L^3\zeta}\sum_{k=1}^{\infty}\frac{(-1)^k(\varepsilon+\sigma
h)}{D^2k^4\pi^4/L^4+4(\varepsilon+\sigma
h)^2})\sin(\theta_l-\theta_r)+\right.\\&&\left.\frac{D\sin(\theta_d-\theta_r)}{LW\zeta^2(\varepsilon+\sigma
h)}+\frac{D\sin(\theta_u-\theta_r)}{LW\zeta^2(\varepsilon+\sigma
h)}\right\}
\end{eqnarray}
and also at the $W=0, L$ terminals:
\begin{eqnarray}\label{eq:I_y0}
&&\nonumber
\frac{I_y(y=0)}{I_{0}}=\int_{-\infty}^{\infty}\frac{d\varepsilon}{\Delta_0}\frac{\Delta^2\tanh(\beta\varepsilon)}{\Delta^2-\varepsilon^2}\sum_{\sigma=\pm}\left\{(\frac{LD}{W^3\zeta^2(\varepsilon+\sigma
h)}\right.\\&&\nonumber\left.+\frac{8LD}{W^3\zeta}\sum_{l=1}^{\infty}\frac{(-1)^l(\varepsilon+\sigma
h)}{D^2l^4\pi^4/W^4+4(\varepsilon+\sigma h)^2})\sin(\theta_
d-\theta_u)+\right.\\&&\left.\frac{D\sin(\theta_d-\theta_r)}{LW\zeta^2(\varepsilon+\sigma
h)}+\frac{D\sin(\theta_d-\theta_l)}{LW\zeta^2(\varepsilon+\sigma
h)}\right\}
\end{eqnarray}
\begin{eqnarray}\label{eq:I_yW}
&&\nonumber
\frac{I_y(y=W)}{I_{0}}=\int_{-\infty}^{\infty}\frac{d\varepsilon}{\Delta_0}\frac{\Delta^2\tanh(\beta\varepsilon)}{\Delta^2-\varepsilon^2}\sum_{\sigma=\pm}\left\{(\frac{LD}{W^3\zeta^2(\varepsilon+\sigma
h)}\right.\\&&\nonumber\left.+\frac{8LD}{W^3\zeta}\sum_{l=1}^{\infty}\frac{(-1)^l(\varepsilon+\sigma
h)}{D^2l^4\pi^4/W^4+4(\varepsilon+\sigma
h)^2})\sin(\theta_d-\theta_u)+\right.\\&&\left.\frac{D\sin(\theta_l-\theta_u)}{LW\zeta^2(\varepsilon+\sigma
h)}+\frac{D\sin(\theta_r-\theta_u)}{LW\zeta^2(\varepsilon+\sigma
h)}\right\}
\end{eqnarray}
$\sigma=\pm$ comes from the spin-dependent nature of the
ferromagnetic material which is sandwiched between the four
superconducting terminals. To be more specific, $I_x(x=0)$,
$I_x(x=L)$, $I_y(y=0)$ and $I_y(y=W)$ represent the Josephson
current in the $x$ direction at $x=0, L$ and $y$ direction at $y=0,
W$, respectively. The above currents involve three sinusoidal terms
whose arguments include phase differences of the lead which
supercurrent is being calculated at and the three other terminals.
As expected, the obtained supercurrents show explicitly that this
interfering terms in the $x$ and $y$ directions vanish for large $L$
and $W$, respectively. This fact is also found in ballistic
junctions \cite{cite:malek1,cite:malek2}. In these two limits,
either large $L$ or $W$, the system takes on quasi-one dimensional
features and we recover the well-known standard sinusoidal Josephson
relation for the supercurrent. However, in the opposite regime where
$L\simeq W$, the proximity-induced order parameters from the
superconducting terminals overlap substantially and additional terms
compared to the one dimensional case appear in the expressions for
the supercurrent. As we shall see, the supercurrent can behave
strongly different from one dimensional junctions as a function of
the phase in one superconducting terminal due to this overlap. In
fact, the supercurrent is a function of a superposition of
sinusoidal phase differences between the different superconducting
leads and one may express the supercurrent relations as
$I(x_{i})=\sum_{j}I_{j}\sin(\theta_{i}-\theta{j})$ in weakly coupled
systems
\cite{cite:omel1,cite:omel2,cite:omel3,cite:malek1,cite:malek2}. The
conservation of charge current is also satisfied by the current
relationships namely, Eqs. (\ref{eq:I_x0}), (\ref{eq:I_xL}),
(\ref{eq:I_y0}) and (\ref{eq:I_yW}). It can be verified explicitly
that:
\begin{align}
I_{x}(x=0)+I_{y}(y=0)=I_{x}(x=L)+I_{y}(y=W).
\end{align}
which constitutes the Kirchhoff law of electricity. We will proceed
to investigate and justify the obtained analytical supercurrent
\textit{numerically} and study how they depend on the
superconducting phases of the terminals. First, we compare our
analytical expressions for the supercurrent with the results
obtained via a macroscopic Ginzburg-Landau theory in the next
subsection.

\subsection{Ginzburg-Landau approach: analytical macroscopic discussions }\label{subsec:GL}
In this subsection, we make a complementary discussion and examine
qualitatively the quasiclassical findings of the previous subsection
by comparison with a phenomenological Ginzburg-Landau (GL) theory
\cite{cite:GL}. The phenomenological approach is a macroscopic
theory which is unable to explain the microscopic mechanism
underlying superconductivity, but instead describes the macroscopic
properties near a phase transition of the system by writing the free
energy as an expansion in the order parameter. We note that the
smallness of the superconducting order parameter may be compared
directly with the weak proximity effect regime in the quasiclassical
theory for temperatures near $T_c$. We assume here that the normal
regions characteristic length scale ($d$) satisfies $\xi\gg d$
where $\xi$ is the coherence
length. In this case the condensation wavefunctions overlap
effectively via the proximity effect. It is instructive to briefly
consider first the one dimensional case, where one may write an
ansatz for the wavefunction as follows
\cite{cite:likharev,cite:larkin}:
\begin{align}
\psi = \psi_1\e{\i\theta_1}\mathcal{X} +
\psi_2\e{\i\theta_2}(1-\mathcal{X}).
\end{align}
Here, $\psi_j$ is the amplitude of the condensate wavefunction in
region $j=1,2$ while $\theta_j$ is the corresponding superconducting
phase. The function $\mathcal{X}$ is unknown, but assumed to satisfy
$\mathcal{X}\to1$ inside region 1 while $\mathcal{X}\to0$ inside
region 2. We now generalize this ansatz to the present four-terminal
two dimensional case. Assume that deep inside the superconducting
banks the order parameter is given as
\begin{equation}\label{eq:psi_GL}
   \psi=\psi_ue^{i\theta_u},\;\psi_de^{i\theta_d},\;\psi_le^{i\theta_l},\;\psi_re^{i\theta_r}.
\end{equation}
Inside the contact region, the four condensation's wavefunctions
overlap and consequently we expect a solution as
\begin{eqnarray}
\nonumber\psi=\psi_{r}e^{i\theta_r}\mathcal{X}\mathcal{Y}(1-\mathcal{Y})
+\psi_le^{i\theta_l}(1-\mathcal{X})\mathcal{Y}(1-\mathcal{Y})+\\\psi_ue^{i\theta_u}\mathcal{Y}\mathcal{X}(1-\mathcal{X})+\psi_de^{i\theta_d}(1-\mathcal{Y})\mathcal{X}(1-\mathcal{X}),
\end{eqnarray}
here we have generalized the mentioned one dimensional ansatz for
the four-terminal junction. The functions $\mathcal{X}$ and
$\mathcal{Y}$ satisfy the following asymptotic behavior:
$\mathcal{X}\rightarrow 0$ in the left, $\mathcal{X}\rightarrow 1$
in the right, $\mathcal{Y}\rightarrow 0$ in the bottom and
$\mathcal{Y}\rightarrow 1$ in the top superconductors. The
supercurrent density can now be defined by the second GL equation
\cite{cite:larkin,cite:likharev}
\begin{equation}\label{GL_supercurrent}
    \mathbf{j}_s=\frac{\alpha\hbar e}{\beta
    m}\text{Im}\left\{\psi^{\ast}\nabla\psi\right\},
\end{equation}
where $\alpha$ and $\beta$ are phenomenological coefficients in the
GL theory. After some calculations, we find the following
expressions for $\mathfrak{j}_x$ and $\mathfrak{j}_y$, the
supercurrent components in the $x$ and $y$ directions,
\begin{eqnarray}\label{GL_jx}
    \nonumber \mathfrak{j}_x&&=\mathcal{X}' (1-\mathcal{Y}) \mathcal{Y} \left\{-\mathcal{Y}
   (1-\mathcal{Y})
   \psi_l\psi_r \sin (\theta_l-\theta_r)-\right.\\&&\nonumber\left.\mathcal{X}^2 (1-\mathcal{Y})
   \psi_d \psi_r \sin
   (\theta_d-\theta_r)-\mathcal{X}^2 \mathcal{Y}
   \psi_u \psi_r\sin (\theta_u-\theta_r)\right.\\&&\nonumber\left.+(1-\mathcal{X})^2 (1-\mathcal{Y})
   \psi_d\psi_l \sin (\theta_d-\theta_l)+\right.\\&&\left.\mathcal{Y}
   (1-\mathcal{X})^2  \psi_u\psi_l \sin (\theta_u-\theta_l)\right\}\\
   \nonumber \mathfrak{j}_y&&= \mathcal{Y}'(1-\mathcal{X}) \mathcal{X} \left\{
   -\mathcal{X}(1-\mathcal{X})  \psi_u\psi_d \sin (\theta_u-\theta_d)-\right.\\&&\nonumber\left.\mathcal{Y}^2(1-\mathcal{X})  \psi_l\psi_u \sin
   (\theta_l-\theta_u)-\mathcal{Y}^2\mathcal{X}
    \psi_r\psi_u \sin
   (\theta_r-\theta_u)\right.\\&&\nonumber\left.+ (1-\mathcal{Y})^2(1-\mathcal{X})
   \psi_l \psi_d \sin
   (\theta_l-\theta_d)+\right.\\&&\left.\mathcal{X} (1-\mathcal{Y})^2 \psi_d \psi_r \sin
   (\theta_r-\theta_d)\right\}
\end{eqnarray}
in which the prime sign denotes derivation. The obtained results
illustrate that, for instance in $\mathfrak{j}_x$, the terms
coupling the top and bottom superconducting terminals vanish. In
this way, we see that the phenomenological GL approach produces
identical dependencies on the superconducting phase differences as
the microscopic approach using quasiclassical theory. Direct
comparison with \eg Eqs. (\ref{eq:I_x0}) and (\ref{eq:I_xL}) in the
appropriate limits for $\mathcal{X}$ shows consistency with Eq.
(\ref{GL_jx}).

\section{Four terminal non-magnetic Josephson junction}\label{sec:SNS_external_field}
In this section, we first set $h=0$ (the exchange field of ferromagnetic
layer) and consider an S/N/S junction. Basically,
there are two methods for inducing a supercurrent into our Josephson
system: 1) via an external flux where the external magnetic field
penetrates the junction through a SQUID-like geometry and 2) via a
current-bias where the supercurrent is injected into the system. A
combination of these two methods is also possible by utilizing
different configurations of a multi-terminal system (for a
comprehensive investigation of such possibilities, see Refs.
\onlinecite{cite:omel1}, \onlinecite{cite:omel2},
\onlinecite{cite:omel3}, \onlinecite{cite:malek1},
\onlinecite{cite:malek2}). The supercurrent at each terminal can be
generally expressed as
$I_i=\sum_{i,j}I_{i,j}\sin(\theta_i-\theta_j)$. Thus if one is able
to tune the superconducting phases independently, the supercurrent
will be a $2\pi$-periodic function of one of the superconducting
phases.

\subsection{Numerical justification of current phase relationships}\label{subsec:numerical_justifications}
In this subsection, we discuss the analytical findings obtained in
the previous section and present numerical results using a real
energy representation. In the actual plots, we consider a
temperature $T=0.05T_c$ and also set the normal region's length and
width to $L=W\simeq 2.5\xi_S$. In this representation, we normalize
lengths against $\xi_S$ and introduce the Thouless energy
$\varepsilon_T=(\hbar D/L^2)$. Also, we have normalized the
quasiparticles' energy by the superconducting gap at zero
temperature $\Delta_0$ and consider units so that $\hbar=k_B=1$.
Moreover, we add a small imaginary number $\eta/\Delta_0=0.1$ to the
quasiparticle energy to account for inelastic scattering which leads
to a finite lifetime for quasiparticle excitations. Setting
$\zeta=7$ ensures the validity of weak proximity in numerical
calculations. Solving numerically the resultant complex boundary
value partial differential equations, the approximate solution
components of the Usadel equation are assumed to be linear
combinations of bicubic Hermite basis functions, and required to
satisfy the Usadel equations (\ref{eq:Linearized Usadel Eq.1}) and
(\ref{eq:Linearized Usadel Eq.2}) exactly at 4 collocation points in
each subrectangle of a grid, and to satisfy the boundary conditions
exactly at certain boundary collocation points. We mention in
passing that we include first-order terms of the anomalous Green's
function in the Kupryianov-Lukichev boundary conditions, as done in
Ref. \onlinecite{cite:iver}, in contrast to the usual approximation
in the literature where such terms are discarded. By doing so, we
improve the accuracy of the analytical solution in our numerical
investigations. Finally, the linear algebraic equations resulting
from the collocation method, which are highly nonsymmetric and thus
difficult to solve using iterative and sparse direct solvers, are
solved using a ``Jacobi" conjugate-gradient method, which means that
the conjugate gradient method (Section 4.8 of Ref.
\onlinecite{cite:sewell2a}) is applied to the preconditioned
equations $D^{-1}A^TA{\bf x} = D^{-1}A^T{\bf b}$, where D is the
diagonal part of $A^TA$.  For a generalized discussion see Ref.
\onlinecite{cite:sewell2b}. The same framework was very recently
used in Ref. \onlinecite{cite:alidoust} to study the anomalous
Fraunhofer pattern appearing in an inhomogeneous S/F/S structure.

\begin{figure}[t!]
\includegraphics[width=8.6cm,height=7.5cm]{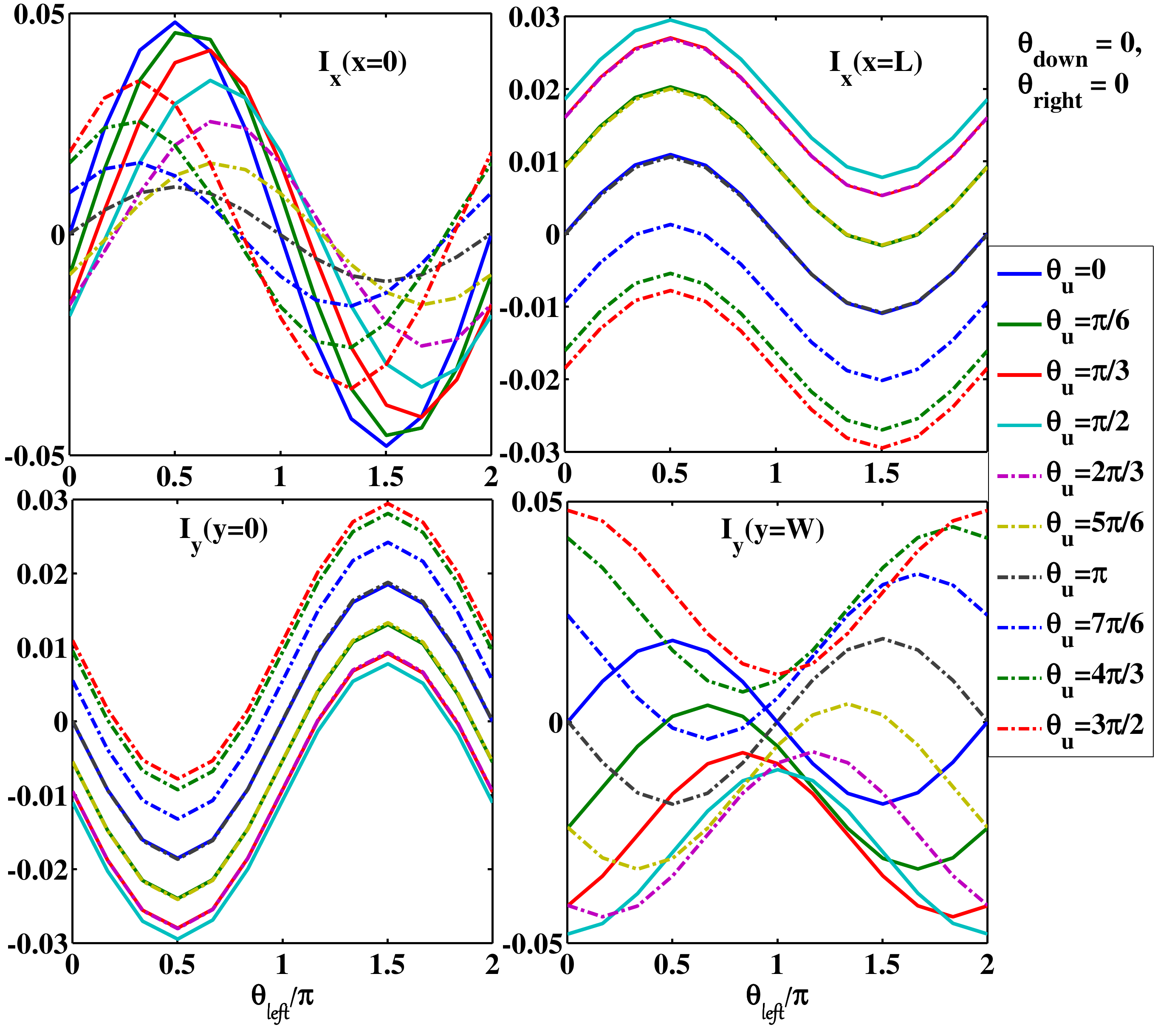}
\caption{\label{fig:thetad_0} \textit{Top left}: Supercurrent in the
$x$ direction as a function of left condensation phase,
$\theta_{\text{left}}$, at left superconductor gate \ie $x=0$.
\textit{Top right}: Supercurrent in the $x$ direction vs left
superconducting phase ,$\theta_{\text{left}}$, at right
superconductor gate \ie $x=L$. \textit{Bottom left}: Supercurrent in
the $y$ direction as a function of left condensation phase at down
superconductor gate \ie $y=0$. \textit{Bottom right}: Supercurrent
in the $y$ direction vs left superconducting phase at up gate \ie
$y=W$. Here other superconductor phases namely, $\theta_{\text{up}}$
and $\theta_{\text{down}}$ are assumed to be zero.}
\end{figure}

In order to clarify the behavior of the supercurrent in the present
four-terminal Josephson junction with respect to condensate phases
of the four superconductors, we use the following strategy. We focus
on the behavior of the supercurrent with respect to one superconductor's
phase (the left one) and set two phases equal to zero:
$\theta_{\text{down}}=\theta_{\text{right}}=0$, while varying
$\theta_{\text{up}}$. The motivation for this is to see if the
supercurrent flowing in one direction can be tuned explicitly by the
superconducting phase difference in the transverse direction, which
would correspond to a superconducting phase transistor-like device.

In general, the supercurrent inside the normal diffusive region is
described by a vector field and depends on the position. The total
flowing current is conserved, as we have proven analytically. We
focus here on the supercurrent flowing into and out of the
terminals, \ie at the positions $x=0$, $y=0$, $x=L$ and $y=W$ gates.
The results are shown in Fig. \ref{fig:thetad_0} where we plot the
supercurrent at the four gates as a function of left superconducting
phase where $\theta_{u}$ is varied while $\theta_{d}=\theta_{r}=0$.
The top left frame shows the supercurrent at $x=0$ as a function of
the left superconducting phase, top right is the supercurrent at
$x=L$, bottom left frame displays the supercurrent at $y=0$, and
finally the bottom right frame shows the supercurrent at $y=W$. The
standard sinusoidal current-phase relation appears at all gates in
the special case where $\theta_{u}$ is equal to zero. This behavior
can be understood by considering Eqs. (\ref{eq:I_x0}),
(\ref{eq:I_xL}), (\ref{eq:I_y0}) and (\ref{eq:I_yW}). In this case,
only terms with $\sin(\theta_l)$ survive and the supercurrent
exhibits a pure sinusoidal relation vs $\theta_l$. When $\theta_{u}$
increases, the phase shift effectively adds a constant which can be
either positive or negative. In particular, the currents at $x=L$
and $y=0$ shift either upwards or downwards depending on the value
of $\theta_u$, as can be understood by looking at Eqs.
(\ref{eq:I_xL}) and (\ref{eq:I_y0}): a change in $\theta_{u}$ only
varies constant terms involving $\sin(\theta_{u})$.

In contrast, variation in $\theta_u$ influences the currents at
$x=0$ and $y=W$ in a more complicated manner. In this case, there is
an explicit dependence on the phase difference $\theta_l-\theta_u$,
which induces a strongly non-sinusoidal behavior in the
current-phase relation. Interestingly, we see
that it is possible to cancel out the current even for a finite
value of $\theta_l$ by choosing $\theta_u$ appropriately. This
observation suggests that the present four-terminal device can act
as a superconducting phase transistor where the phase difference in
one direction controls the supercurrent flowing in the perpendicular
direction. The underlying mechanism behind this is the interference
between the condensate wavefunctions in the diffusive normal region,
which results in an intricate phase-dependence of the supercurrent
as shown in the analytical results.

\section{Four-terminal magnetic Josephson junction}\label{sec:SFS}

In this section, we consider a four-terminal Josephson junction with
a ferromagnetic barrier where the exchange field of the magnetic
layer is oriented along the $z$ direction. In the usual two-terminal
magnetic Josephson junctions, an increment of the ferromagnetic
barrier thickness not only reverses the current direction at
particular thicknesses but also renders the minimum of junction
energy to change from $0$ superconducting phase difference to a
$\pi$ phase. The phenomenon is so called 0-$\pi$ transition. As has
been discussed in Ref. \onlinecite{cite:malek1} the junction energy
where there are several superconducting leads can be expressed as
$E_J=\sum_{j< i}\gamma_{j,i}(1-\cos(\theta_j-\theta_i))$. Here, the
$i$ and $j$ indices stand for the $i$th and $j$th superconducting
leads. Below, we demonstrate that an increment in the thickness of
the ferromagnet can reverse the flow of supercurrent into a pair of
the superconducting terminals (along the direction of increment),
whereas the current direction in the other terminal pair remains
unaltered.

\subsection{The behavior of critical supercurrent as a function of magnetic barrier thickness}\label{subsec:0-pi}

\begin{figure}[t!]
\includegraphics[width=8.6cm,height=7.1cm]{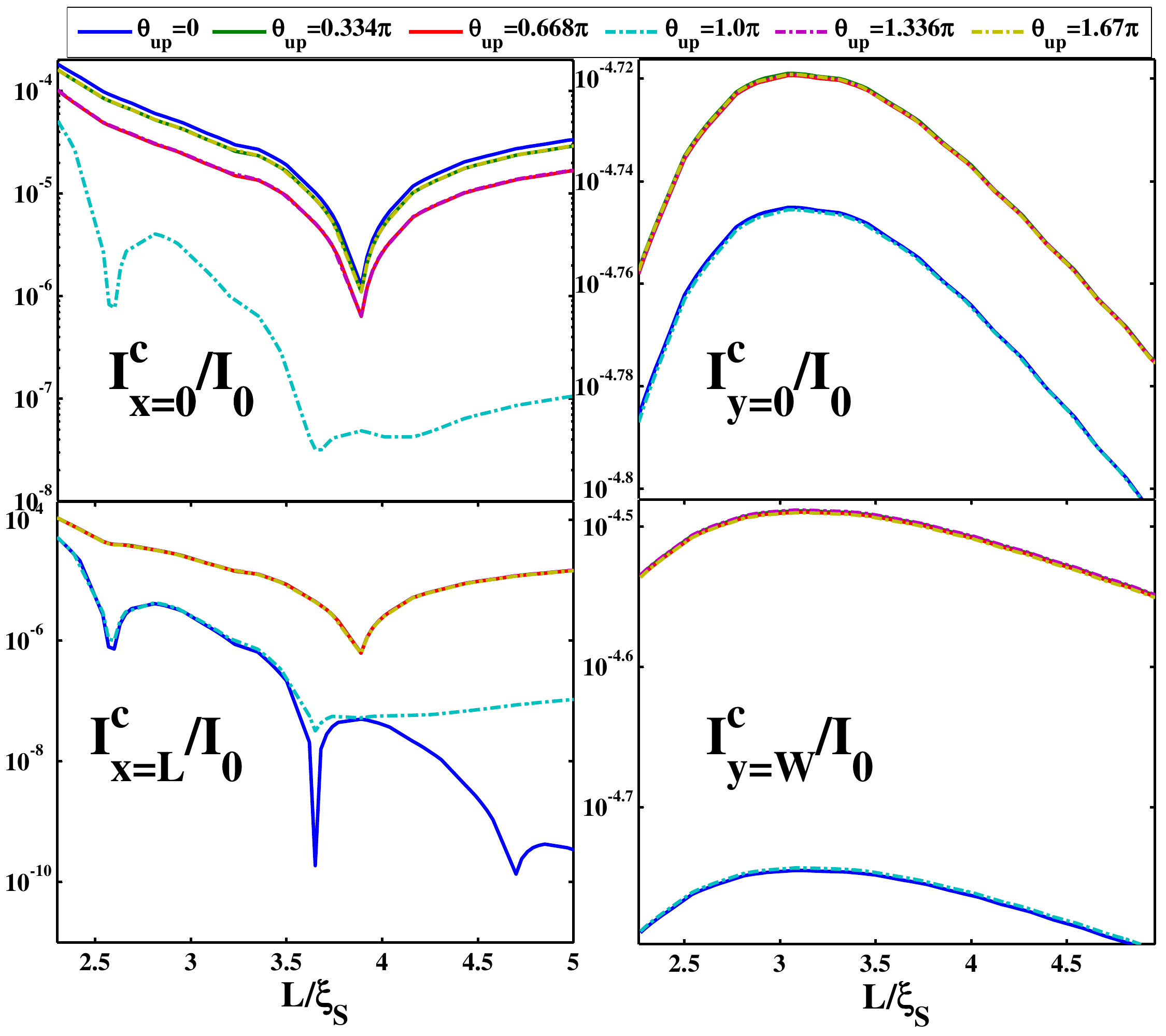}
\caption{\label{fig:SFS} Critical supercurrent as a function
of the normalized junction length $L/\xi_S $ at different
superconducting gates and for various values of $\theta_{up}$, the
superconducting phase of the up terminal. \textit{Top left}: at the left
superconductor gate \ie $x=0$. \textit{Top right}: at the right
superconductor gate \ie $x=L$. \textit{Bottom left}: at the down
superconductor gate \ie $y=0$. \textit{Bottom right}: at the up gate \ie
$y=W$. The other superconductor phases are fixed at zero.}
\end{figure}

We here present a \textit{numerical} study of the transport
properties of four-terminal ferromagnetic Josephson junctions.
Although the numerical results are confirmed by the analytical
expressions presented in Sec. \ref{sec:SNS_external_field}, we
include first-order terms of the anomalous Green's function in the
Kupryianov-Lukichev boundary conditions in contrast to the
approximation used for deriving the analytical expressions for
supercurrent where such terms are dropped. We now consider a
non-zero value of the ferromagnetic exchange field $h$. For a weak,
diffusive ferromagnetic alloy such as
$\text{Pd}_{x}\text{Ni}_{1-x}$, the exchange field $h/\Delta_0$ is
tunable by means of the doping level $x$ to take values in the range
meV to tens of meV. Here, we will fix $h=10\Delta_0$, which
typically places the exchange field $h$ in the range $10$-$20$ meV.
In order to investigate the effects of magnetic barrier thickness on
the supercurrent at each terminal and the influence of the various
superconducting phases, we follow a similar strategy as in the
previous section. $\theta_{l}$ is varied from $0$ to $2\pi$ where
magnetic barrier length, $L$, is being varied from $L=2\xi_S$ to
$L=5\xi_S$. The other superconducting phases are fixed at zero
except $\theta_{u}$ which is changed in order to demonstrate the
possible influence of the other superconducting phases. The critical
value of the supercurrent at each terminal is calculated separately
for each value of $\theta_u$.

Fig. \ref{fig:SFS} indicates the behavior of critical supercurrent
at each superconductor lead as a function of normalized junction
length $L/\xi_S$ for various values of $\theta_{u}$. The top left
frame exhibits the critical current at left terminal. Except for
$\theta_{u}=\pi$ which shows two points changing the supercurrent
direction, the other values give rise to one sign-change in the
critical current. Identical qualitative behavior appears for the
current at the right terminal except when $\theta_{u}=0$, as shown
in the bottom left frame. Top and bottom right frames exhibit the
critical supercurrent vs $L/\xi_S$ at the down and up terminals,
respectively. The critical supercurrent at the two terminals show a
smooth function of $L/\xi_S$ which is in stark contrast with the
behavior of the critical supercurrent at the left and right
terminal. Thus, the increment of the junction length
primarily affects the critical supercurrent flowing into leads along the same
direction of the increment. Moreover, the direction of the current can be drastically switched
by tuning the superconducting phase of up terminal. In contrast, the
current flowing into the superconducting banks perpendicular
direction to junction length increment is left unchanged. This class of multi-terminal
ferromagnet Josephson junction then offers an interesting synthesis
between 0 and $\pi$-states, and possibly $\phi$-states, due to the
fact that the coefficients $I_j$ can change sign depending on the
junction parameters such as $L$ and $W$.

\section{Conclusions}\label{sec:Conclusions}
In conclusion, we have studied a four-terminal Josephson junction
where a diffusive normal or ferromagnetic metal with sides $L$ and
$W$ is sandwiched among four $s$-wave superconductor leads. We have
obtained explicit analytical results using the quasiclassical
Keldysh-Usadel method for the supercurrent in the system. We find
that the wavefunctions of the four superconductors interfere
efficiently when $L\simeq W$ and modifies the standard sinusoidal
current-phase relation which confirm previous findings in ballistic
junctions. These findings are confirmed qualitatively by using a
macroscopic Ginzburg-Landau theory. We have presented numerical
results for the behavior of the supercurrent, and demonstrated that
the current flowing along one axis may be tuned by the
superconducting phase-difference along the perpendicular direction.
It is demonstrated that such four-terminal junctions can provide a
rich switching circuit element (due to additional degrees of freedom
in comparison with one-dimensional two-terminal Josephson junctions)
where the various superconducting phases influence considerably the
current behavior at the terminals. In particular, we show that a
reversal in critical current direction as a function of junction
length can be strongly switched by means of variation of
superconducting phase of perpendicular terminals. The present
investigations of diffusive cruciate magnetic Josephson junction may
provide new perspectives for the design of a superconducting phase
switches where can be used in quantum circuits as switching
elements.

\textit{\textbf{\underline{Acknowledgments:}}}  We would like to
thank K. Halterman for his generosity regarding compiler source and
also F. S. Bergeret for fruitful discussions.

\end{document}